\documentclass[aps,preprint,nofootinbib,floatfix]{revtex4}%
\usepackage{amssymb}
\usepackage{amsmath}
\usepackage{epsfig}
\usepackage{amsfonts}
\usepackage{graphicx}%
\providecommand{\U}[1]{\protect\rule{.1in}{.1in}}
\begin{document}
\title{Kummer function and High energy String Scatterings}
\author{Sheng-Lan Ko}
\email{slko.py96g@g2.nctu.edu.tw}
\affiliation{Department of Electrophysics, National Chiao-Tung University, Hsinchu, Taiwan,
R.O.C. }
\author{Jen-Chi Lee}
\email{jcclee@cc.nctu.edu.tw}
\affiliation{Department of Electrophysics, National Chiao-Tung University and Physics
Division, National Center for Theoretical Sciences, Hsinchu, Taiwan, R.O.C.}
\affiliation{Institute of Physics, Academia Sinica, Taipei, Taiwan, R.O.C.}
\author{Yi Yang}
\email{yiyang@mail.nctu.edu.tw}
\affiliation{Department of Electrophysics, National Chiao-Tung University and Physics
Division, National Center for Theoretical Sciences, Hsinchu, Taiwan, R.O.C.}
\date{\today }

\begin{abstract}
Based on a summation algorithm for Stirling number identity developed
recently, we discover that the ratios calculated previously among high energy
string scattering amplitudes in the Gross regime (GR) can be extracted from
the Kummer function of the second kind. This function naturally shows up in
the leading order of high energy string scattering amplitudes in the Regge
regime (RR). As a result, the identity suggested by string theory calculation
can be rigorously proved by a totally different but sophisticated mathematical
method. We conjecture and give evidences that the existence of these ratios in
the RR persists to all orders in the Regge expansion of high energy string
scattering amplitudes.

\end{abstract}
\maketitle

High energy limits of scattering amplitudes are of fundamental importance in
quantum mechanics, quantum field theory and string theory. Not only can they
be used to greatly simplify a lot of mathematical calculations of the
amplitudes but also that one can use the high energy amplitudes to extract
many fundamental characteristics of the physical theory. There are two
fundamental regimes of high energy scattering amplitudes, namely, the fixed
angle regime and the fixed momentum transfer regime. These two regimes
represent two different high energy perturbation expansions of the scattering
amplitudes, and contain complementary information of the underlying theory. In
QCD, for example, the probe of high energy, fixed angle regime reveals the
partonic structures of hadrons, quarks and gluons. On the other hand, the
Regge behavior of high energy hadronic scattering amplitudes suggested a
string model of hardons with a linear relation between hadron spins and their
mass squared. In string theory, the scattering amplitudes in the high energy,
fixed angle regime \cite{GM, Gross, GrossManes}, the Gross regime (GR), were
recently intensively reinvestigated for massive string states at arbitrary
mass levels \cite{ChanLee1,ChanLee2, CHL,CHLTY,PRL,susy,Closed}. See also
\cite{West1,West2,Moore}. An infinite number of linear relations, or stringy
symmetries, among string scattering amplitudes of different string states were
obtained. Moreover, these linear relations can be solved for each fixed mass
level $M^{2}=2(n-1)$, and ratios $T^{(n,2m,q)}/T^{(n,0,0)},n\geq
2m+2q;m,q\geq0$ among the amplitudes can be obtained. An important new
ingredient of these calculations is the decoupling of zero-norm states (ZNS)
\cite{ZNS1,ZNS3,ZNS2} in the old covariant first quantized (OCFQ) string
spectrum. Since so far there does not exist any algebraic structure (or group
structure) of this high energy 26D spacetime symmetry (except $\omega_{\infty
}$ for the case of toy 2D string theory \cite{ZNS3}), mathematically the
meaning of these infinite number of ratios remains mysterious.

\bigskip

In this letter, we calculate high energy massive string scattering amplitudes
in the fixed momentum transfer regime, the Regge regime (RR). There have been
some previous studies \cite{RR1,RR2,RR3,RR4,RR5,RR6} of high energy string
scatterings in this regime\ in the literature. Our motivation here is to
calculate massive string scatterings and try to extract possible pattens of
the scattering amplitudes in the RR which may mimic the patterns (e.g. the
ratios mentioned above) in the GR. Since the decoupling of ZNS applies to all
kinematic regime, it is reasonable to expect some implication of this
decoupling in the RR. We found that \cite{KLY} the number of high energy
scattering amplitudes for each fixed mass level in the RR is much more
numerous than that of GR calculated previously. On the other hand, it seems
that both the saddle-point method and the method of decoupling of high energy
ZNS adopted in the calculation of GR do not apply to the case of RR. However
the calculation is still manageable, and the general formula for the high
energy scattering amplitudes for each fixed mass level in the RR can be
written down explicitly \cite{KLY}. As expected, there is no linear relation
anymore as in the case of scatterings in the GR. Moreover, we discover that
the leading order amplitudes at each fixed mass level in the RR can be
expressed in terms of the Kummer function of the second kind. More
surprisingly, for those leading order high energy amplitudes $A^{(n,2m,q)}$ in
the RR with the same type of $(n,2m,q)$ as those of GR, we can extract from
them the above mentioned ratios $T^{(n,2m,q)}/T^{(n,0,0)}$ in the GR by using
Kummer function of the second kind, which naturally shows up in the leading
order of high energy string scattering amplitudes in the RR. The calculation
brings a link between high energy string scattering amplitudes in the GR and
the RR. In addition to the decoupling of ZNS calculated previously
\cite{ChanLee1,ChanLee2, CHL,CHLTY,PRL,susy,Closed} (or the unitarity of
quantum string theory), the calculation from Kummer function in this letter
seems to give another interpretation of \ the existence of these ratios.
Finally, we calculate some Regge string scattering amplitudes to subleading
orders and conjecture that these ratios persist to all orders in the Regge
expansion of high energy string scattering amplitudes.

We stress that, mathematically, the proof of the identification of the ratios
in the GR from the Kummer function calculated in the RR turns out to be highly
nontrivial. This is based on a summation algorithm for Stirling number
identity derived by Mkauers in 2007 \cite{MK}. It is very interesting to see
that the identity in Eq.(\ref{14}) suggested by string theory calculation can
be rigorously proved by a totally different but sophisticated mathematical
method. Although this kind of coincidence is not unusual in the development of
string theory, our results bring an interesting connection between string
theory and combinatoric number theory. Moreover, the connection between Kummer
function and high energy string scatterings may shed light on a deeper
understanding of stringy symmetries.

We begin with a brief review of high energy string scatterings in the GR. That
is in the kinematic regime $s,-t\rightarrow\infty$, $t/s\approx-\sin^{2}%
\frac{\theta}{2}$= fixed (but $\theta\neq0$) where $s,t$ and $u$ are the
Mandelstam variables and $\theta$ is the CM scattering angle. It was shown
\cite{CHLTY,PRL} that for the 26D open bosonic string the only states that
will survive the high-energy limit at mass level $M_{2}^{2}=2(n-1)$ are of the
form
\begin{equation}
\left\vert n,2m,q\right\rangle \equiv(\alpha_{-1}^{T})^{n-2m-2q}(\alpha
_{-1}^{L})^{2m}(\alpha_{-2}^{L})^{q}\left\vert 0,k_{2}\right\rangle \label{1}%
\end{equation}
where the polarizations of the 2nd particle with momentum $k_{2}$ on the
scattering plane were defined to be $e^{P}=\frac{1}{M_{2}}(E_{2}%
,\mathrm{k}_{2},0)=\frac{k_{2}}{M_{2}}$ as the momentum polarization,
$e^{L}=\frac{1}{M_{2}}(\mathrm{k}_{2},E_{2},0)$ the longitudinal polarization
and $e^{T}=(0,0,1)$ the transverse polarization. Note that $e^{P}$ approaches
to $e^{L}$ in the GR, and the scattering plane is defined by the spatial
components of $e^{L}$ and $e^{T}$. Polarizations perpendicular to the
scattering plane are ignored because they are kinematically suppressed for
four point scatterings in the high-energy limit. One can then use the
saddle-point method to calculate the high energy scattering amplitudes. For
simplicity, we choose $k_{1}$, $k_{3}$ and $k_{4}$ to be tachyons and the
final result of the ratios of high energy, fixed angle string scattering
amplitude are \cite{CHLTY,PRL}%
\begin{equation}
\frac{T^{(n,2m,q)}}{T^{(n,0,0)}}=\left(  -\frac{1}{M_{2}}\right)
^{2m+q}\left(  \frac{1}{2}\right)  ^{m+q}(2m-1)!!. \label{2}%
\end{equation}
The ratios in Eq.(\ref{2}) can also be obtained by using the decoupling of two
types of ZNS in the spectrum. As an example, for $M_{2}^{2}=4$ we get
\cite{ChanLee1,ChanLee2}
\begin{equation}
T_{TTT}:T_{LLT}:T_{(LT)}:T_{[LT]}=8:1:-1:-1. \label{3}%
\end{equation}
To convince the readers that the infinite ratios in Eq.(\ref{2}) are the
symmetries or, at least, remnant of full spacetime symmetries of 26D string
theory, it was shown that a set of 2D discrete ZNS $\Omega_{J_{1},M_{1}}^{+}%
$carry $\omega_{\infty}$symmetry charges \cite{ZNS3}%
\begin{equation}
\int\frac{dz}{2\pi i}\Omega_{J_{1},M_{1}}^{+}(z)\Omega_{J_{2},M_{2}}%
^{+}(0)=(J_{2}M_{1}-J_{1}M_{2})\Omega_{(J_{1}+J_{2}-1),(M_{1}+M_{2})}^{+}(0).
\label{26}%
\end{equation}

A natural question arises. Is there any mathematical structure (e.g. group
structure) of these infinite number of ratios? Let's consider a simple analogy
from partical physics. The ratios of the nucleon-nucleon scattering processes%
\begin{align}
(a)\text{ \ }p+p  &  \rightarrow d+\pi^{+},\nonumber\\
(b)\text{ \ }p+n  &  \rightarrow d+\pi^{0},\nonumber\\
(c)\text{ \ }n+n  &  \rightarrow d+\pi^{-}%
\end{align}
can be calculated to be%
\begin{equation}
T_{a}:T_{b}:T_{c}=1:\frac{1}{\sqrt{2}}:1 \label{27}%
\end{equation}
from $SU(2)$ isospin symmetry. Similarly, as we will see in the rest of the
paper, the ratios in Eq.(\ref{2}) can be extracted from Kummer function. The
key is to study high energy string scatterings in the RR.

We now turn to the discussion on high energy string scatterings in the RR.
That is in the kinematic regime $s\rightarrow\infty$, $\sqrt{-t}$= fixed (but
$t\neq-\infty$). It can be shown \cite{KLY} that the number of high energy
scattering amplitudes for each fixed mass level in the RR is much more
numerous than those calculated from Eq.(\ref{1}) in the GR. For our purpose
here, however, we will only calculate scattering amplitudes corresponding to
the vertex in Eq.(\ref{1}). The relevant kinematics are%
\begin{equation}
e^{P}\cdot k_{1}\simeq-\frac{s}{2M_{2}},\text{ \ }e^{P}\cdot k_{3}\simeq
-\frac{\tilde{t}}{2M_{2}}=-\frac{t-M_{2}^{2}-M_{3}^{2}}{2M_{2}}; \label{4}%
\end{equation}%
\begin{equation}
e^{L}\cdot k_{1}\simeq-\frac{s}{2M_{2}},\text{ \ }e^{L}\cdot k_{3}\simeq
-\frac{\tilde{t}^{\prime}}{2M_{2}}=-\frac{t+M_{2}^{2}-M_{3}^{2}}{2M_{2}};
\label{5}%
\end{equation}
and%
\begin{equation}
e^{T}\cdot k_{1}=0\text{, \ \ }e^{T}\cdot k_{3}\simeq-\sqrt{-{t}}. \label{6}%
\end{equation}
Note that $e^{P}$ \textit{does not} approach to $e^{L}$ in the RR. The Regge
scattering amplitude for the $s-t$ channel can be calculated to be%
\begin{align}
A^{(n,2m,q)}  &  =\int_{0}^{1}dx\,x^{k_{1}\cdot k_{2}}(1-x)^{k_{2}\cdot k_{3}%
}\left[  \frac{e^{T}\cdot k_{3}}{1-x}\right]  ^{n-2m-2q}\nonumber\\
&  \left[  \frac{e^{L}\cdot k_{1}}{-x}+\frac{e^{L}\cdot k_{3}}{1-x}\right]
^{2m}\left[  \frac{e^{L}\cdot k_{1}}{x^{2}}+\frac{e^{L}\cdot k_{3}}{(1-x)^{2}%
}\right]  ^{q}\nonumber\\
&  \simeq(\sqrt{-{t}})^{n-2m-2q}\left(  \frac{\tilde{t}^{\prime}}{2M_{2}%
}\right)  ^{q}\int_{0}^{1}dx\,x^{k_{1}\cdot k_{2}}(1-x)^{k_{2}\cdot
k_{3}-n+2m}\nonumber\\
&  \sum_{j=0}^{2m}{\binom{2m}{j}}\left(  \frac{s}{2M_{2}x}\right)  ^{j}\left(
\frac{-\tilde{t}^{\prime}}{2M_{2}(1-x)}\right)  ^{2m-j}\nonumber\\
&  =(\sqrt{-{t}})^{n-2m-2q}\left(  \frac{\tilde{t}^{\prime}}{2M_{2}}\right)
^{q}\left(  \frac{\tilde{t}^{\prime}}{2M_{2}}\right)  ^{2m}\nonumber\\
&  \sum_{j=0}^{2m}{\binom{2m}{j}}(-1)^{j}\left(  \frac{s}{\tilde{t}^{\prime}%
}\right)  ^{j}B\left(  k_{1}\cdot k_{2}-j+1,k_{2}\cdot k_{3}-n+j+1\right)  .
\label{7}%
\end{align}
Note that the term $\frac{e^{L}\cdot k_{1}}{x^{2}}$ in the bracket is
subleading in energy and can be neglected. In the high energy limit, the beta
function in Eq.(\ref{7}) can be approximated by%
\begin{equation}
B\left(  k_{1}\cdot k_{2}-j+1,k_{2}\cdot k_{3}-n+j+1\right)  \simeq B\left(
-1-\frac{1}{2}s,-1-\frac{t}{2}\right)  \left(  -\frac{s}{2}\right)
^{-j}\left(  -1-\frac{t}{2}\right)  _{j} \label{8}%
\end{equation}
where $(a)_{j}=a(a+1)(a+2)...(a+j-1)$ is the Pochhammer symbol. Finally, the
leading order amplitude in the RR can be written as%
\begin{align}
A^{(n,2m,q)}  &  =B\left(  -1-\frac{s}{2},-1-\frac{t}{2}\right)  \sqrt
{-t}^{n-2m-2q}\left(  \frac{1}{2M_{2}}\right)  ^{2m+q}\nonumber\\
&  2^{2m}(\tilde{t}^{\prime})^{q}U\left(  -2m\,,\,\frac{t}{2}+2-2m\,,\,\frac
{\tilde{t}^{\prime}}{2}\right)  , \label{9}%
\end{align}
which is UV power-law behaved as expected. $U$ in Eq.(\ref{9}) is the Kummer
function of the second kind and is defined to be%
\begin{equation}
U(a,c,x)=\frac{\pi}{\sin\pi c}\left[  \frac{M(a,c,x)}{(a-c)!(c-1)!}%
-\frac{x^{1-c}M(a+1-c,2-c,x)}{(a-1)!(1-c)!}\right]  \text{ \ }(c\neq2,3,4...)
\label{10}%
\end{equation}
where $M(a,c,x)=\sum_{j=0}^{\infty}\frac{(a)_{j}}{(c)_{j}}\frac{x^{j}}{j!}$ is
the Kummer function of the first kind. $U$ and $M$ are the two solutions of
the Kummer equation%
\begin{equation}
xy^{^{\prime\prime}}(x)+(c-x)y^{\prime}(x)-ay(x)=0. \label{11}%
\end{equation}
At this point, it is crucial to note that, in our case of Eq.(\ref{9}),
$c=\frac{t}{2}+2-2m$ and is not a constant as in the usual definition, so $U$
in Eq.(\ref{9}) is\textit{ not} a solution of the Kummer equation. This will
make our follow-up analysis, the proof of Eq.(\ref{14}) discussed below, more
complicated as we will see soon. On the contrary, since $a=-2m$ an integer,
the Kummer function in Eq.(\ref{9}) terminated to be a finite sum (see
Eq.(\ref{12}) below). This will simplify the manipulation of Kummer function
used in this paper. We stress that all the calculations in this paper do not
rely on Kummer equation Eq.(\ref{11}). In fact, one can take Eq.(\ref{12})
below as a formal definition of Kummer function used in this paper.

It is important to note that there is no linear relation among high energy
string scattering amplitudes of different string states for each fixed mass
level in the RR as can be seen from Eq.(\ref{9}). This is very different from
the result in the GR in Eq.(\ref{2}). In other words, the ratios
$A^{(n,2m,q)}/A^{(n,0,0)}$ are $t$-dependent functions. In particular, we can
extract the coefficients of the highest power of $t$ in $A^{(n,2m,q)}%
/A^{(n,0,0)}$. We can use the identity of the Kummer function%
\begin{align}
&  2^{2m}(\tilde{t}^{\prime})^{-2m}\ U\left(  -2m,\frac{t}{2}+2-2m,\frac
{\tilde{t}^{\prime}}{2}\right)  =\,_{2}F_{0}\left(  -2m,-1-\frac{t}{2}%
,-\frac{2}{\tilde{t}^{\prime}}\right) \nonumber\\
&  \equiv\sum_{j=0}^{2m}\left(  -2m\right)  _{j}\left(  -1-\frac{t}{2}\right)
_{j}\frac{\left(  -\frac{2}{\tilde{t}^{\prime}}\right)  ^{j}}{j!}=\sum
_{j=0}^{2m}{\binom{2m}{j}}\left(  -1-\frac{t}{2}\right)  _{j}\left(  \frac
{2}{\tilde{t}^{\prime}}\right)  ^{j} \label{12}%
\end{align}
to calculate
\begin{equation}
\frac{A^{(n,2m,q)}}{A^{(n,0,0)}}=(-1)^{q}\left(  \frac{1}{2M_{2}}\right)
^{2m+q}(-t)^{m}\sum_{j=0}^{2m}(-2m)_{j}\left(  -1-\frac{t}{2}\right)
_{j}\frac{(-2/t)^{j}}{j!}+\mathit{O}\left\{  \left(  \frac{1}{t}\right)
^{m+1}\right\}  \label{13}%
\end{equation}
where we have replaced $\tilde{t}^{\prime}$ by $t$ as $t$ is large. \textit{If
the leading order coefficients in Eq.(\ref{13}) extracted from the high energy
string scattering amplitudes in the RR are to be identified with the ratios
calculated previously among high energy string scattering amplitudes in the GR
in Eq.(\ref{2}), we need the following identity}
\begin{align}
&  \sum_{j=0}^{2m}(-2m)_{j}\left(  -1-\frac{t}{2}\right)  _{j}\frac
{(-2/t)^{j}}{j!}\nonumber\\
&  =0(-t)^{0}+0(-t)^{-1}+...+0(-t)^{-m+1}+\frac{(2m)!}{m!}(-t)^{-m}%
+\mathit{O}\left\{  \left(  \frac{1}{t}\right)  ^{m+1}\right\}  . \label{14}%
\end{align}
The coefficient of the term $\mathit{O}\left\{  \left(  1/t\right)
^{m+1}\right\}  $ in Eq.(\ref{14}) is irrelevant for our discussion. The proof
of Eq.(\ref{14}) turns out to be nontrivial. The standard approach by using
integral representation of the Kummer function seems not applicable here.
Presumably, the difficulty of the rigorous proof of Eq.(\ref{14}) is
associated with the unusual non-constant $c$ in the argument of Kummer
function in Eqs.(\ref{9}) and (\ref{12}) as mentioned above. It is a
nontrivial task to do the proof compared to the usual cases where the argument
$c$ of the Kummer function is a constant. Here we will adopt another approach
to prove Eq.(\ref{14}). This approach strongly relies on the algorithm for
Stirling number identity derived by Mkauers \cite{MK} in 2007, and is highly
nontrivial either. The leading order identity of Eq.(\ref{14}) can be written
as
\begin{equation}
f(m)\equiv\sum_{j=0}^{m}(-1)^{j}{\binom{2m}{j+m}}\left[
s(j+m-1,j-1)+s(j+m-1,j)\right]  =(2m-1)!! \label{15}%
\end{equation}
where the signed first Stirling number $s(n,k)$ is defined to be
\begin{equation}
(x)_{n}=\sum_{k=0}^{n}(-1)^{n-k}s(n,k)x^{k}. \label{16}%
\end{equation}
The authors had verified the validity of Eq.(\ref{15}) for $m=1,2...,2000$
before they carried out the exact proof to be discussed below. To prove
Eq.(\ref{15}) we define
\begin{equation}
f(u,m)\equiv\sum_{j=0}^{m+u}(-1)^{j}{\binom{2m+u}{j+m}}\left[
s(j+m-1,j-1)+s(j+m-1,j)\right]  \label{17}%
\end{equation}
with $f(0,m)=f(m)$. By using the result of \cite{MK}, one can prove that
$f(u,m)$ satisfies the following recurrence relation
\begin{equation}
-(1+2m+u)f(u,m)+(2m+u)f(u+1,m)+f(u,m+1)=0. \label{18}%
\end{equation}
Eq.(\ref{18}) is the most nontrivial step to prove Eq.(\ref{15}). Finally by
taking $u=0$, it can be shown that the second term of Eq.(\ref{18}) vanishes
\cite{KLY}. Eq.(\ref{15}) is then proved by mathematical induction. The
vanishing of the coefficients of $(-t)^{0},(-t)^{-1},...(-t)^{-m+1}$ terms on
the LHS of Eq.(\ref{14}) means, for $1\leqslant i\leqslant m$,
\begin{equation}
g(m,i)\equiv\sum_{j=0}^{m+i}(-1)^{j-i}{\binom{2m}{j+m-i}}\left[
s(j+m-1-i,j)+s(j+m-1-i,j-1)\right]  =0. \label{19}%
\end{equation}
To prove this identity, we need the recurrence relation \cite{MK}
\begin{align}
-  &  2(1+m)^{2}(1+2m)g(m,i)+(2+7m+4m^{2})g(m+1,i)\nonumber\\
-  &  2m(1+m)(1+2m)g(m+1,i+1)-m\times g(m+2,i)=0. \label{20}%
\end{align}
Putting $i=0,1,2..$, and using the fact we have just proved, i.e.
$g(m+1,0)=(2m+1)g(m,0)$, one can prove Eq.(\ref{19}). Eq.(\ref{14}) is thus
finally proved. It is very interesting to see that the identity in
Eq.(\ref{14}) suggested by string scattering amplitude calculation can be
rigorously proved by a totally different but sophisticated mathematical
method. In conclusion, ratios in Eq.(\ref{2}) can be extracted from Kummer
function of the second kind%
\begin{equation}
\frac{T^{(n,2m,q)}}{T^{(n,0,0)}}=\left(  -\frac{1}{2M}\right)  ^{2m+q}%
2^{2m}\lim_{t\rightarrow\infty}(-t)^{-m}U\left(  -2m\,,\,\frac{t}%
{2}+2-2m\,,\,\frac{t}{2}\right)  . \label{21}%
\end{equation}
In view of Eq.(\ref{27}), this result may help to uncover the fundamental
symmetry of string theory. At last, we give an explicit calculation of the
high energy string scattering amplitudes to subleading orders in the RR for
$M_{2}^{2}=4$ \cite{KLY}
\begin{equation}
A_{TTT}\sim\frac{1}{8}\sqrt{-t}ts^{3}+\frac{3}{16}\sqrt{-t}t(t+6)s^{2}%
+\frac{3t^{3}+84t^{2}-68t-864}{64}\sqrt{-t}\,s+O(1), \label{22}%
\end{equation}%
\begin{align}
A_{LLT}  &  \sim\frac{1}{64}\sqrt{-t}(t-6)s^{3}+\frac{3}{128}\sqrt{-t}%
(t^{2}-20t-12)s^{2}\nonumber\\
&  \quad\quad+\frac{3t^{3}-342t^{2}-92t+5016+1728(-t)^{-1/2}}{512}\sqrt
{-t}\,s+O(1), \label{23}%
\end{align}%
\begin{align}
A_{(LT)}  &  \sim-\frac{1}{64}\sqrt{-t}(t+10)s^{3}-\frac{1}{128}\sqrt
{-t}(3t^{2}+52t+60)s^{2}\nonumber\\
&  \quad\quad-\frac{3[t^{3}+30t^{2}+76t-1080-960(-t)^{-1/2}]}{512}\sqrt
{-t}\,s+O(1), \label{24}%
\end{align}%
\begin{align}
A_{[LT]}  &  \sim-\frac{1}{64}\sqrt{-t}(t+2)s^{3}-\frac{3}{128}\sqrt
{-t}(t+2)^{2}s^{2}\nonumber\\
&  \quad\quad-\frac{(3t-8)(t+6)^{2}[1-2(-t)^{-1/2}]}{512}\sqrt{-t}\,s+O(1).
\label{25}%
\end{align}
We have ignored an overall irrelevant factors in the above amplitudes. Note
that the calculation of Eq.(\ref{24}) and Eq.(\ref{25}) involves amplitude of
the state $(\alpha_{-2}^{T})(\alpha_{-1}^{L})\left\vert 0,k_{2}\right\rangle $
which can be shown to be of leading order in the RR \cite{KLY}, but is of
subleading order in the GR as it is not in the form of Eq.(\ref{1}). However,
the contribution of the amplitude calculated from this state will not affect
the ratios $8:1:-1:-1$ in the RR \cite{KLY}. One can now easily see that the
ratios of the coefficients of the highest power of $t$ in these leading order
coefficient functions $\frac{1}{8}:\frac{1}{64}:-\frac{1}{64}:-\frac{1}{64}$
agree with the ratios in the GR calculated in Eq.(\ref{3}) as expected.
Moreover, one further obeservation is that these ratios remain the same for
the coefficients of the highest power of $t$ in the subleading orders
$(s^{2})$ $\frac{3}{16}:\frac{3}{128}:-\frac{3}{128}:-\frac{3}{128}$ and $(s)$
$\frac{3}{64}:\frac{3}{512}:-\frac{3}{512}:-\frac{3}{512}$. More examples will
be given in \cite{KLY}. We thus conjecture that the existence of these GR
ratios of Eq.(\ref{2}) in the RR persists to all orders in the Regge expansion
of high energy string scattering amplitudes.

In conclusion, physically, the connection between Kummer function and high
energy string scattering amplitudes derived in this letter may shed light on a
deeper understanding of stringy symmetries. In contrast to an infinite number
of ratios obtained in the GR previously, here one gets a nice Kummer function
in the RR, which surely contains more analytic properties than just ratios for
string scatterings. Mathematically, the proof of identity in Eq.(\ref{14})
brings an interesting bridge between string theory and combinatoric number
theory. Finally, in addition to the Kummer function in the leading order
amplitudes discoverd in this paper, the structure of the coefficient functions
of the subleading order amplitudes in the RR maybe the most exciting
mathematical problems to study.

Acknowledgement

This work is supported in part by the National Science Council, 50 billions
project of MOE and National Center for Theoretical Science, Taiwan. We
appreciated the correspondence of Dr. Manuel Mkauers at RISC, Austria for his
kind help of providing us with the rigorous proof of Eq.(\ref{15}). J.C. would
like to thanks the support and the hospitality of Institute of Physics at
Academia Sinica where part of this work is finalized. He is indebted to
H.Y.Cheng, S.P. Li, H.N. Li, H.L.Yu and T.C.Yuan for many of their enlighting discussions.

\end{document}